\documentclass[10pt]{article}
\usepackage{epsf,latexsym,amssymb}
\usepackage[mathscr]{eucal} %see p.8 of AMSfonts guide; w/o option, redefined
\newcommand{\comment}[1]{}
\newcommand{\edo}{\end{document}}

\usepackage{amsmath}  %will give errors with \be \ee !!!
\newtheorem{theorem}{Theorem}
\newtheorem{itlemma}{Lemma}[section] %number by section (set in \em by default)
\newtheorem{itproposition}[itlemma]{Proposition}
\newtheorem{itcorollary}[itlemma]{Corollary}
\newtheorem{itremark}[itlemma]{Remark}
\newtheorem{itdefinition}[itlemma]{Definition}
\newtheorem{itexample}[itlemma]{Example}

\newenvironment{lemma}{\begin{itlemma}\rm}{\end{itlemma}} %no-italics
\newenvironment{remark}{\begin{itremark}\rm}{\end{itremark}} %no-italics
\newenvironment{corollary}{\begin{itcorollary}\rm}{\end{itcorollary}}
\newenvironment{proposition}{\begin{itproposition}\rm}{\end{itproposition}}
\newenvironment{definition}{\begin{itdefinition}\rm}{\end{itdefinition}}
\newenvironment{example}{\begin{itexample}\rm}{\end{itexample}}

\newcommand{\be}[1]{\begin{equation}\label{#1}}
\newcommand{\ee}{\end{equation}}
\newcommand{\bl}[1]{\begin{lemma}\label{#1}}
\newcommand{\ble}[1]{\begin{lemmaex}\label{#1}}
\newcommand{\br}[1]{\begin{remark}\label{#1}}
\newcommand{\bt}[1]{\begin{theorem}\label{#1}}
\newcommand{\bd}[1]{\begin{definition}\label{#1}}
\newcommand{\bp}[1]{\begin{proposition}\label{#1}}
\newcommand{\bc}[1]{\begin{corollary}\label{#1}}
\newcommand{\bfact}[1]{\begin{fact}\label{#1}}
\newcommand{\ber}[1]{\begin{exercise}\label{#1}}
\newcommand{\bex}[1]{\begin{example}\label{#1}}
\newcommand{\bem}[1]{\begin{example}\label{#1}}  %Yes, 2 different ones...
\newcommand{\ec}{\mybox\end{corollary}}
\newcommand{\efact}{\mybox\end{fact}}
\newcommand{\eer}{\mybox\end{exercise}}
\newcommand{\eex}{\mybox\end{example}}
\newcommand{\eem}{\mybox\end{example}}
\newcommand{\el}{\mybox\end{lemma}}
\newcommand{\ele}{\mybox\end{lemmaex}}
\newcommand{\er}{\mybox\end{remark}}
\newcommand{\et}{\qed\end{theorem}}
\newcommand{\ed}{\mybox\end{definition}}
\newcommand{\ep}{\mybox\end{proposition}}
\newcommand{\epr}{\end{proof}}
\newcommand{\bpr}{\begin{proof}}

\newcommand{\ecs}{\end{corollary}}
\newcommand{\eers}{\end{exercise}}
\newcommand{\eexs}{\end{example}}
\newcommand{\eems}{\end{example}}
\newcommand{\els}{\end{lemma}}
\newcommand{\eles}{\end{lemmaex}}
\newcommand{\ers}{\end{remark}}
\newcommand{\ets}{\end{theorem}}
\newcommand{\eds}{\end{definition}}
\newcommand{\eps}{\end{proposition}}
\newcommand{\halmos}{\rule{1ex}{1.4ex}}
\newcommand{\qed}{\hfill \halmos} %put \qed at right margin
\newcommand{\mybox}{\hfill $\Box$} %put \qed at right margin (white square)
\newcommand{\beq}{\begin{eqnarray}}
\newcommand{\eeq}{\end{eqnarray}}
\newcommand{\beqn}{\begin{eqnarray*}}
\newcommand{\eeqn}{\end{eqnarray*}}
\newcommand{\bi}{\begin{itemize}}
\newcommand{\ei}{\end{itemize}}
\newcommand{\ben}{\begin{enumerate}}
\newcommand{\een}{\end{enumerate}}

\newenvironment{proof}{\noindent {\em Proof}.\ }{\hspace*{\fill}$\halmos$\medskip}

\newcommand{\bs}{\begin{split}}
\newcommand{\es}{\end{split}}

\title{Network reconstruction based on quasi-steady state data}
\author{Eduardo D.\ Sontag\\
Department of Mathematics, Rutgers University}

\usepackage{graphicx}
\newcommand{\pic}[2]{\includegraphics[scale=#1]{#2}}
\newcommand{\picc}[2]{\centerline{\pic{#1}{#2}}}

\newcommand{\barp}{{\bar p}}
\newcommand{\barx}{{\bar x}}
\newcommand{\bart}{{\bar t}}
\newcommand{\pbar}{\bar p}

\begin{document}

\maketitle

\centerline{Abstract}

\medskip

\noindent
This note discusses a theoretical issue regarding the application of the
``Modular Response Analysis'' method to quasi-steady state (rather than
steady-state) data.

\medskip

\section{Introduction}

%% START pasting from intro to NYACAD pape:

The reverse engineering problem in systems biology is, loosely speaking, that
of unraveling the web of interactions among the components of protein and gene
regulatory networks.
A major goal is to map out the direct functional interactions among
components, a problem that is difficult to approach by means of standard
statistical and machine learning  approaches such as clustering into
co-expression patterns. 
Information on direct functional interactions throws light upon the
possible mechanisms and architecture underlying the observed behavior of
complex molecular networks.

An intrinsic difficulty in capturing direct interactions in intact cells by
traditional genetic experiments, RNA interference, hormones, growth factors,
or pharmacological interventions, is that any perturbation to a particular
gene or signaling component may rapidly propagate throughout the network, thus
causing \emph{global} changes which cannot be easily distinguished from direct
(\emph{local}) effects.
Thus, one central goal in reverse engineering
is to use the observed global responses
(such as steady-state changes in concentrations of active
proteins, mRNA levels, or transcription rates)
in order to infer the local interactions between individual nodes.
One potentially very powerful approach to solve the global to local problem is
the Modular Response Analysis (MRA) or ``unraveling'' method proposed
in
\cite{kholodenkoPNAS02}
%\cite{arXiv}
and further elaborated upon in
\cite{SontagKiyatinKholodenko04,
kholodenkoJTB05,
berman-dasgupta-sontag-nyacad,
berman_dasgupta_sontag}
(see
\cite{Stark_at_al_2003,Crampin_et_al_2004}
for reviews).

The MRA experimental design compares the steady states that result after
performing independent perturbations to each
``modular component'' of a network.  These perturbations might be
genetic or biochemical, or (in eukaryotes) they might be achieved
through the down-regulation of protein levels by means of RNAi.
This latter experimental approach to MRA was the one taken in
\cite{Santos_et_al_2007},
which quantified positive and negative feedback effects in the Raf/Mek/Erk
MAPK network in rat adrenal pheochromocytoma (PC-12) cells.  
Using the formulas given in
\cite{SontagKiyatinKholodenko04}
and
\cite{kholodenkoJTB05},
%kholodenkoJTB05},
the authors of \cite{Santos_et_al_2007}
uncovered connectivity differences depending on whether the cells are
stimulated with epidermal growth factor (EGF) or instead with neuronal growth
factor (NGF).
There are a couple of subtle theoretical gaps, however, when applying MRA
algorithms to data like that employed in \cite{Santos_et_al_2007}.
The main gap is due to the fact that the data fed into the MRA algorithm
included non-steady state measurements.
Specifically, for EGF stimulation,
network responses were measured at the peak of Erk activity (at 5 minutes)
%(and 5 min NGF for comparison) 
and \emph{not at steady state}.
This note fills that gap, providing a theoretical justification for the use of
quasi-steady state information.

\subsection{Mathematical formulation}

% again pasting from NYACAD:

We assume that there are $n$ quantities $x_i(t)$ that can be in principle
measured, such as the levels of activity of selected proteins, or
the transcription rates of certain genes.
These quantities are thought of as state variables in a dynamical system
and are collected into a time-dependent vector $x(t)=(x_1(t),\ldots ,x_n(t))$.
The dynamical system is described by a system of differential
equations:
\beqn
\dot x_1 &=& f_1(x_1,\ldots ,x_n,p_1,\ldots ,p_m) \\
\dot x_2 &=& f_2(x_1,\ldots ,x_n,p_1,\ldots ,p_m) \\
        &\vdots&\\
\dot x_n     &=& f_n(x_1,\ldots ,x_n,p_1,\ldots ,p_m)
\eeqn
or, in more convenient vector form,
\[
\dot x=f(x,p)
\]
(dot indicates time derivative, and arguments $t$ are omitted when clear).
The $p_i$'s are parameters, collected into a vector $p=(p_1,\ldots ,p_m)$.
These parameters can be manipulated, but, once changed, they remain constant
for the duration of the experiment. 
An example would be that in which the variables $x_i$ correspond to the levels
of protein products coresponding to $n$ genes in a network, and the parameters
reflect translation rates, controlled by RNAi.  Another example would be the
total 
levels of proteins whose half-lives are long compared to the time scale of the
processes being described by the differential equations, such as
phosphorylation modifications of these proteins in a signaling pathway.

The ultimate goal is to obtain, for each pair of variables $x_i$ and $x_j$,
the relative signs and magnitudes of the partial derivatives
\[
\frac{\partial f_i}{\partial x_j}\,,
\]
which quantify the \emph{direct} effects of each variable $x_j$ on each
variable $x_i$.

The critical assumption for MRA, and indeed the main point
of~\cite{kholodenkoPNAS02,arXiv,SontagKiyatinKholodenko04}, is that,
while one may not know the detailed form of the vector field $f$,
often one does know which parameters $p_j$ directly affect which
variables $x_i$.
For example, $x_i$ may be the level of activity of a particular protein,
and $p_i$ might be the total amount (active plus inactive) of that 
particular protein; in that case, we know that $p_i$ only directly affects
$x_i$.

%% end adapted from NYACAD

In the standard version of MRA, one first measures a steady state
$\barx$ corresponding to a ``wild type'' vector of parameters $\barp$,
that is $f(\barx,\barp)=0$.
Subsequent perturbations are separately performed to each entry of $\barp$,
and a new steady state is measured, one for each such perturbation.
Using these data (and assuming a that certain independence condition
which we review later is satisfied), it is possible to calculate,
at least in the ideal noise-free case,
the Jacobian of $f$, evaluated at $(\barx,\barp)$, up to
a scalar multiplicative factor uncertainty on each row.
(Such uncertainty is unavoidable when using only steady state measurements,
since multiplying a row of the vector field $f$ by a nonzero constant does not
affect the location of steady states.)
A variation of MRA is possible, which allows for the use of non-steady state,
time-series data.  However, this alternative method, developed in
\cite{SontagKiyatinKholodenko04},
requires one to compute time derivatives, and hence is hard to apply when
time measurements are spaced far apart and/or are noisy.
%%%
An intermediate possibility is to use quasi-steady state data, meaning, just
as in the experimental setup of \cite{Santos_et_al_2007}, that one employs
data collected at times when a variable has been observed to attain a local
maximum or local minimum.  That is the case addressed in this note.

More precisely, we will consider the following scenario.  For any fixed
variable, let us say the $i$th component $x_i$ of
$x$, we consider some time instant $\bart_i$ at which $\dot x_i(t)$ is zero.
%This includes the cases of a time point at which $x_i$ has been observed to
%attain a local maximum, or a local minumim, or (as in the more classical
%version of MRA) a time large enough that $x$ may be considered to be in steady
%state.
Under the same independence hypothesis as in the classical MRA case, plus
the nondegeneracy assumption
%, in the non-steady state case, 
that the second time derivative $\ddot x_i(\bart_i)$ is not zero (so that we
have a true local minimum or local maximum, but not an inflection point), we
show here that the MRA approach applies in exactly the same manner as in the
steady-state case. 
Specifically, the $i$th row of the Jacobian of $f$, evaluated at the vector
$(\barx,\barp)$, is recovered up to a constant multiple, where
$\barx=x(\bart_i)$ is the full state $x$ at time $\bart_i$.
The main difference with the steady-state case is that different rows of $f$
are estimated at different pairs $(\barx,\barp)$, since the considered times
$\bart_i$ at which each individual $\dot x_i(t)$ vanishes are in general
different for different indices $i$, and so the state $\barx$ is different for
different $i$'s.

\section{Using quasi-steady state data}

We fix an index $i\in \{1,\ldots ,n\}$, and an initial condition $x(0)$,
and assume that the solution $x(t)$ with this initial condition and
a given parameter vector $\barp$ has the property that, for some time
$\bart=\bart_i$, we have that both
$\dot x_i(\bart)=0$ and 
$\ddot x_i(\bart)\not= 0$.
At the instant $t=\bart$, $x_i$ achieves a local minimum or a local
maximum as a function of $t$.
We describe the reconstruction of the $i$th row of the Jacobian of $f$, which
is the same as the gradient $\nabla f_i$, where $f_i$ is the $i$th coordinate
of $f$, evaluated at $x=\barx$ and $p=\barp$, where $\barx=x(\bart)$.

To emphasize the dependence of the solution on the parameters (the initial
condition $x(0)$ will remain fixed), we will denote the solution
of the differential equation $\dot x=f(x,p)$ by $x(t,p)$.
The function $x(t,p)$ is jointly continuously differentiable in $x$ and $p$,
if the vector field $f$ is continuously differentiable (see e.g.~\cite{mct},
Appendix C).
Note that, with this notation, the left-hand side of the differential
equation can also be written as $\partial x/\partial t$, and that
$x(\bart,\barp)=\barx$.

We introduce the following function:
\[
\alpha (t,p) = \frac{\partial x_i}{\partial t}(t,p) = f_i(x(t,p),p).
\]
Note that $\alpha (\bart,\barp)=0$.  Also,
\[
\frac{\partial \alpha }{\partial t} (t,p) \;=\;
\frac{\partial^2 x_i}{\partial t^2}(t,p) \;=\;
\nabla f_i(x(t,p),p)\,f(x(t,p),p).
\]
The assumption that $\ddot x_i(\bart)\not= 0$ when $p=\barp$ means
that $\frac{\partial \alpha }{\partial t} (\bart,\barp)\not= 0$.
Therefore, we may apply the implicit function theorem and conclude the
existence of a mapping $\tau $, defined on a neighborhood of $\barp$,
with the property that
\[
\alpha (\tau (p),p) = 0
\quad\quad\mbox{for all}\;\; p\approx \barp
\]
and $\tau (\barp)=\bart$ (and, in fact, $t=\tau (p)$
is the unique value of $t$ near $\bart$ such that
$(\partial x_i/\partial t)(t,p)=\alpha (t,p)=0$).

We define, also in a neighborhood of $\barp$, the differentiable function
\[
\varphi(p) = x(\tau (p),p)
\]
and note that $\varphi(\barp)=\barx$.
Observe that, from the definition of $\alpha $, we have:
\be{maineqn}
f_i(\varphi(p),p)=0   \quad\quad\mbox{for all}\;\; p\approx \barp .
\ee

We next discuss how to reconstruct $\nabla f_i(\barx,\barp)$, up to a constant
multiple, under the assumption (as in \cite{kholodenkoPNAS02}) that it is
possible to apply $n-1$ independent parameter perturbations to all species
different from the $i$th one.
This discussion is basically identical to that for the steady state
case, given in
\cite{kholodenkoPNAS02,kholodenkoJTB05,berman-dasgupta-sontag-nyacad}.

Mathematically, we assume that there are $n-1$ indices
$j_1,j_2,\ldots ,j_{n-1}$ with the properties that
(a) $f_i$ does not depend directly on any $p_j$:
$\partial f_i/\partial p_j\equiv 0$, for $j\in \{j_1,j_2,\ldots ,j_{n-1}\}$,
and
(b) the vectors $v_j=(\partial \varphi/\partial p_j)(\barp)$, for these $j$'s, are
linearly independent. 
Assumption (a) is structural, and is key to the method and nontrivial,
but assumption (b) is a weak genericity assumption.

We then have, taking total derivatives in~(\ref{maineqn}):
\[
\nabla f_i(\barx,\barp) \, v_j \;=\;0,
\quad
\quad
j \in  \{p_{j_1},p_{j_2},\ldots ,p_{j_{n-1}}\}.
\]
Thus, the vector $\nabla f_i(\barx,\barp)$ is orthogonal to the
$n-1$ dimensional subspace spanned by $\{v_1,\ldots ,v_{n-1}\}$,
and hence is uniquely determined up to multiplication by a positive scalar.
Another way to phrase this is to say that $\nabla f_i(\barx,\barp)$ is in the
(one-dimensional) left nullspace of the matrix $A$ whose rows are the $v_i$'s,
or that (if nonzero) the transpose of this gradient can be found as an (any)
eigenvector associated to the zero eigenvalue of the transpose of $A$.

\subsubsection*{Numerical approximation by finite differences}

Approximating the vectors $v_j$ by finite differences, one has that
$\nabla f_i(\barx,\barp)$ is approximately orthogonal to these differences
as well.
Explicitly, suppose that we approximate
$v_j=(\partial (\varphi/\partial p_j)(\barp)$ by:
\[
\frac{1}{h} \left(\varphi(\barp + he_j) - \barx\right),
\]
where $h$ is small and where $e_j$ is the vector having a one in the $j$th
position and zeros elsewhere.
Then, $\nabla f_i(\barx,\barp)$ is (approximately) orthogonal to
the differences
\[
\varphi(\barp + he_j) - \barx,
\]
which form a set of $n-1$ linearly independent vectors (if $h$ is small).
A simple matrix inversion (after fixing an arbitrary value for one of its
entries) allows the computation of $\nabla f_i(\barx,\barp)$.
Observe that division by the potentially small number $h$ is not required in
performing these operations,  In fact, no knowledge whatsoever about
parameter values is needed by the algorithm.

Note that $\varphi(\barp + he_j)$ is the state $x(t)$ at the time $t$ at which
\emph{the particular coordinate} $x_i$ achieves a local extremum value, if the
parameters have been perturbed to $p=\barp + he_j$.
To be more precise, $t$ is the unique time close to $\bart$ such that
$\dot x_i(t)=0$ when parameter vector $p$ is being used.
Theoretically, we must have $p\approx\barp$, so
$h$ must be very small, but, in practice, quite large perturbations of $p$ also
work fine.

\section{A simple numerical example}

\begin{figure}[ht]
\picc{0.5}{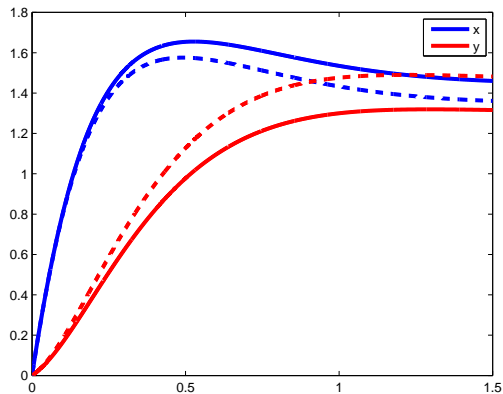}
\caption{Trajectories; dashed is perturbed motion with 25\% change in parameter}
\label{fig1}
\end{figure}

\begin{figure}[ht]
\picc{0.5}{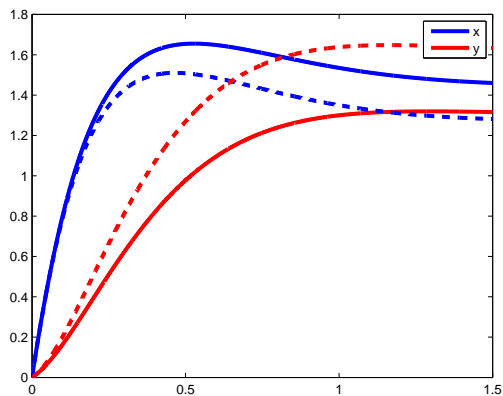}
\caption{Trajectories; dashed is perturbed motion with 50\% change in parameter}
\label{fig2}
\end{figure}

\begin{figure}[ht]
\picc{0.5}{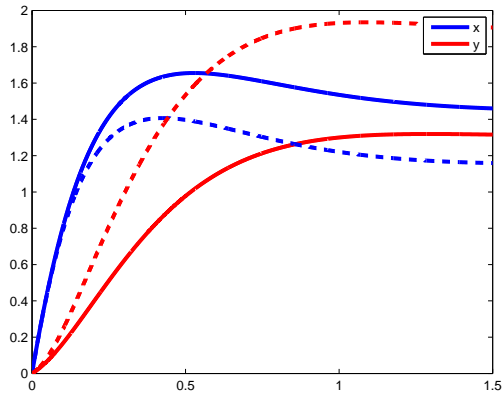}
\caption{Trajectories; dashed is perturbed motion with 100\% change in parameter}
\label{fig3}
\end{figure}

We illustrate the calculations with a very simple example, the following
system (writing $x$ instead of $x_1$ and $y$ instead of $x_2$):
\beqn
\dot x&=&-3x+\frac{10}{1+y}\\
\dot y&=&px + 1 - 3y
\eeqn
with initial state $(0,0)$ and reference parameter $\barp=2$.
This might represent the simplified dynamics of a two-gene network, in which
the first gene enhances the expression of the second gene, which in turn
represses the rate of expression of the first one, 
there is a constitutive rate of production of the second gene, and
both protein products decay at rate 3 sec$^{-1}$.
The single parameter $p$ may represent a promoter strength, and we assume that
there is a way to perturb it (perhaps by duplication or sequence change).
The solid lines in Figure~\ref{fig1} (and also in Figures~\ref{fig2} 
and~\ref{fig3}) show plots of the solution coordinates $x(t)$ and $y(t)$.

Let us pose the following problem: not knowing the above equations, estimate
the relative strength of the second gene's effect on the rate of expression of
the first one.  The only data to be used are the levels of both gene products
($x(t)$ and $y(t)$) at the time when $x(t)$ achieves its local maximum.
We do assume known the fact that the parameter $p$ affects \emph{directly}
only the rate of expression of the second gene, not the first.
Observe that the maximum of $x$ is attained at $t\approx0.5275$, and the
values there are (approximately) $x(t)=1.6553$ and $y(t)=1.0138$.
The gradient $\nabla f_1$ of $-3x+\frac{10}{1+y}$, evaluated at
$(1.407,1.3695)$, has the true (but unknown to the algorithm)
value: 
\[
\left(-3,\left.-\frac{10}{(1+y)^2}\right|_{y=1.0138}\right) \approx (-3,-2.4659).
\]

Next, we perform the ``experiment'' in which $p$ is up-perturbed by 25\%.
With the new parameter $p=2.5$, we obtain plots as shown by the dotted
lines in Figure~\ref{fig1}.  
Now the maximum of $x$ is attained at $t\approx0.4268$, and the
values there are $x(t)=1.407$ and $y(t)=1.3695$.
%(MATLAB implementation: if y denotes the solution vector,
%[m,I] = min(y(:,1)) selects the minimum index I, and y(I,:) is therefore
%the estimate of $\varphi(p)$.)
Letting $\delta =(1.407,1.3695)-(1.6553,1.0138)$, the unknown (to the algorithm)
gradient $\nabla f_1$ is known to be (approximately) orthogonal to $\delta $.  
Any vector perpendicular to $\delta $ must be a multiple of
$(-3,-2.3455)$.  (We normalized the first entry to -3 merely in order to
compare our result to the true gradient; the algorithm does not know the value
``-3''.  In practice, however, one may assume that the first entry of the
vector is negative, reflecting degradation or dilution effects, so the
algorithm will give the correct sign for the second term, as well as its
magnitude relative to the rate of degradation or dilution.)
The relative error in our estimate is less than 5\%.

Even larger perturbations may be performed.  For example, a 50\% perturbation
from $\pbar=2$ to $p=3$, provides the dashed lines in 
Figure~\ref{fig2}.
Now the maximum for $x$ is attained at $t\approx0.4658$, and there
$x(t)=1.5103$ and $y(t)=1.2073$.
The estimated gradient is now
$(-3,-2.2476)$, which gives a relative error of less than 9\%.
Finally, a 100\% perturbation
to $p=4$ provides the dashed lines in 
Figure~\ref{fig3}.
Now the maximum for $x$ is attained at $t\approx0.4268$, and there
$x(t)=1.4071$ and $y(t)=1.3695$.
The estimated gradient is now
$(-3,-2.0936)$, which gives a relative error of about 15\%.

%end comment linear example

%\bibliography{qss-unravel_rev}
%\end{document}

\subsubsection*{Remarks}

As its name implies, one of the main advantages of the MRA method in the
steady-state case is that only ``communicating intermediates'' in-between
``modules'' need to be measured (for example, just the active forms of Erk1/2,
Mek1/2 and Raf-1, in~\cite{Santos_et_al_2007}).
Here, we only carried out the analysis in the case in which all the variables
$x_i$ can be measured.
In the general case, if one assumes that hidden (internal) variables are at
quasi-steady state at the same times as the communicating variables, then
an implicit function argument as in~\cite{arXiv} allows one to reduce
to the present situation, by writing the hidden variables in terms of the
communicating quantities.  However, there is no reason for the
method to work when the hidden variables do not have this property.
% like this: say y is a hidden variable; if g(x(t(p)),y(t(p)),z(t(p)),p)\equiv 0
% then solve for y(t(p)) in terms of x, z, and p, and substitute into the
% equation for x
% note that the property is OK when e.g. there are two forms of a protein and
% there is conservation; it is not a priori true if there are three forms...

Also, we assume perfect ``noise free'' data.  The analysis of noise performed
in~\cite{kholodenkoJTB05} carries over with no changes to the quasi-steady
state case.

\end{document}